\documentclass{aastex63}

\newcommand{\vecr}{\mbox{\boldmath $r$}}

\begin{document}


\title{A Numerical Method for Determining the Elements of Circumbinary
  Orbits and Its Application to Circumbinary Planets and the
  Satellites of Pluto-Charon}
\author{Jason Man Yin Woo}
\altaffiliation{Present address: Institute for Computational Science,
  University of Zurich, Winterthurerstrasse 190, CH-8057 Zurich,
  Switzerland}
\affiliation{Department of Earth Sciences, The University of Hong Kong,
  Pokfulam Road, Hong Kong}
\affiliation{Earth-Life Science Institute, Tokyo Institute of
  Technology, Meguro, Tokyo 152-8550, Japan}
\author[0000-0003-1930-5683]{Man Hoi Lee}
\affiliation{Department of Earth Sciences, The University of Hong Kong,
  Pokfulam Road, Hong Kong}
\affiliation{Department of Physics, The University of Hong Kong,
  Pokfulam Road, Hong Kong}


\begin{abstract}
Planets and satellites orbiting a binary system exist in the solar
system and extrasolar planetary systems.
Their orbits can be significantly different from Keplerian orbits, if
they are close to the binary and the secondary-to-primary mass ratio
is high.
A proper description of a circumbinary orbit is in terms of the free
eccentricity $e_{\rm  free}$ at the epicyclic frequency $\kappa_0$,
forced eccentricity $e_{\rm forced}$ at the mean motion $n_0$,
and oscillations at higher frequencies forced by the non-axisymmetric
components of the binary's potential.
We show that accurate numerical values for the amplitudes and
frequencies of these terms can be extracted from numerical orbit
integrations by applying fast Fourier transformation (FFT) to the
cylindrical distance between the circumbinary object and the center of
mass of the binary as a function of time.
We apply this method to three Kepler circumbinary planets and the
satellites of Pluto-Charon.
For the satellite Styx of Pluto-Charon, the FFT results for $\kappa_0$
and $e_{\rm free}$ differ significantly from the first-order analytic
value and the value reported by \cite{sho15}, respectively.
We show that the deviation in $\kappa_0$ is likely due to the
effect of the 3:1 mean-motion resonance and discuss the implications
of the lower value for $e_{\rm free}$.
\end{abstract}

\section{INTRODUCTION}

Four small satellites -- Styx, Nix, Kerberos and Hydra -- are
orbiting around Pluto-Charon, which is a binary system with Charon
about $1/8$ of the mass of Pluto. Due to the non-spherical potential
of the binary and the proximity of the satellites to the binary (with
the orbital periods of the small satellites $\approx 3.16$--$5.98$
times that of Pluto-Charon),
the orbits of the small satellites are significantly
different from Keplerian orbits \citep{lee06}. To precisely describe a
circumbinary orbit is not an easy task. Using the osculating Keplerian
orbital elements may not be accurate enough, because they are
calculated from the angular momentum and energy assuming a Keplerian
orbit.
Figure \ref{aeosc} shows the variations of the osculating semimajor
axis $a_{\rm osc}$ and eccentricity $e_{\rm osc}$ for the two
satellites closest to Pluto-Charon, Styx and Nix,
using the best fit of \cite{sho15}.
We observe that both $a_{\rm osc}$ and $e_{\rm osc}$ have
non-trivial variations on orbital timescales.
The variation of $a_{\rm osc}$ is $\sim 4\%$ and $2\%$ for Styx and
Nix, respectively.
The percentage variation is calculated by $({\rm maximum} - {\rm
  minimum})/{\rm mean} \times 100\%$, and the mean value is indicated
by the black dashed lines in the upper panels of Figure \ref{aeosc}.
For $e_{\rm osc}$, its lowest value can be more than 10 times smaller
than the largest value within an orbital period. Also, the mean values
of $e_{\rm osc}$, $\langle e_{\rm osc}\rangle$, and the free
eccentricities $e_{\rm free}$ (which are indicated by the black and
red dashed lines, respectively, in the lower panels of Figure
\ref{aeosc}) are significantly different from each other, where $e_{\rm
  free}$ is the appropriate generalization of eccentricity for
circumbinary orbits (see below).
The values of $e_{\rm free}$ shown in Figure \ref{aeosc} are those
determined in Section \ref{sec:pluto} and listed in Table \ref{table2}
below.
Because of the variations in the osculating orbital elements and
the significant difference between $\langle e_{\rm osc}\rangle$ and
$e_{\rm free}$, the osculating elements do not provide clear
information about the orbital evolution history of the system, and it
is not suitable to use the instantaneous or mean value of the
osculating elements to describe a circumbinary orbit.
The orbital parameters reported by \cite{bro15} and \cite{sho15} were
obtained by an alternative method: fitting a precessing ellipse to the
numerical integration of the best-fit orbit or the observational data,
but it has not been demonstrated that the eccentricity obtained in
this way should agree with $e_{\rm free}$.

Besides Pluto-Charon in the solar system, binary stars are frequently
observed in the Milky Way.
A number of exoplanets orbiting around binary stars have been
discovered.
Currently, there are about 20 of them, including Kepler-16~b,
Kepler-34~b, and Kepler-35~b.
Kepler-16~b was the first circumbinary exoplanet discovered by the
{\it Kepler} spacecraft and it is a Saturn-mass planet orbiting around
a binary with a total mass of $0.89 M_{\odot}$, where $M_{\odot}$ is
the solar mass (\citealt{doy11}).
This was quickly followed by the discovery of Kepler-34~b and
Kepler-35~b (\citealt{wel12}).
Kepler-34~b is $0.22 M_{\rm J}$ and orbits a pair of stars with masses
similar to the Sun, while Kepler-35~b is about half the mass of
Kepler-34~b ($0.13 M_{\rm J}$) and orbits a binary with masses $0.89
M_{\odot}$ and $0.81 M_{\odot}$.
One of the more recent discovery is Kepler-1647~b, which is the largest
circumbinary Kepler planet known and has the longest period for any
confirmed transiting circumbinary exoplanets (\citealt{kos16}).
A major difference between Pluto-Charon and the binary stars
with circumbinary planets is that the orbit of the present-day
Pluto-Charon is circular, while the orbits of the binary stars are
eccentric.

Since circumbinary orbits can be significantly non-Keplerian, an
analytic theory has to be developed to
describe them. \cite{lee06} showed that the motion of a test particle
orbiting around a circular binary can be represented by the
superposition of the circular motion of the guiding center, the
epicyclic motion, the forced oscillations due to the non-axisymmetric
components of the binary's potential, and the vertical motion.
This theory is first order in terms of deviations from the
uniform circular motion of the guiding center.
\cite{leu13} extended this theory to describe the orbit of a
test particle around an eccentric binary, to first order in the
eccentricity of the binary, and found additional forced oscillation
terms.

In this paper, we investigate how the amplitudes and frequencies of
the various terms describing the motion of a circumbinary object can
be extracted from numerical orbit integrations and whether the first-order
epicyclic theory is adequate in describing the orbits of the known
circumbinary objects.
In Section \ref{sec:fft}, we present the idea of using fast Fourier
transformation (FFT) to determine the amplitudes and frequencies of
the various terms describing the motion of a circumbinary object.
Then we present in Section \ref{sec:result} the results of FFT on
three Kepler circumbinary planets (Kepler-16~b, Kepler-34~b, and
Kepler-35~b) and the satellites of Pluto-Charon.
For Styx, the FFT results for the epicyclic frequency (and hence the
periapse precession frequency) and $e_{\rm free}$ differ significantly
from the analytic value from the first-order epicyclic theory and the
value reported by \cite{sho15}, respectively.
In Section \ref{sec:theory}, we develop an alternative analytic theory
using the disturbing function approach and show that the deviation in
the epicyclic frequency is due to the modification of the secular
precession frequency by the 3:1 mean-motion resonance.
In Section \ref{sec:con}, our findings are summarized and the
implications for the origin and evolution of the satellites of
Pluto-Charon are discussed.

\section{EPICYCLIC THEORY AND ORBITAL ANALYSIS USING FFT}
\label{sec:fft}

In the first-order epicyclic theory of \cite{lee06} and \cite{leu13},
the circumbinary object is assumed to have negligible mass and is
treated as a test particle.
The orbit of the test particle about the center of mass of the binary
is modelled as small deviations from the circular motion of a guiding
center at $R_0$:
\begin{equation}
R = R_0 + R_1(t),\ \phi = \phi_0(t) + \phi_1(t),\ z = z_1(t) ,
\end{equation}
where $\phi_0(t) = n_0 t + \varphi_0$, $n_0$ is the mean motion, and
$\varphi_0$ is a constant.
\cite{lee06} showed that the orbit around a circular binary can be
represented by the superposition of the circular motion of the guiding
center, the epicyclic motion (with amplitude $e_{\rm free}$ and
epicyclic frequency $\kappa_0$), the forced oscillations due to the
non-axisymmetric components of the binary's potential (with amplitudes
$C_k^0$ and frequencies equal to
$|k (n_0 -n_{AB})|$, where $n_{AB}$ is the mean motion of the binary
and $k = 1$, 2, 3, $\ldots$), and the vertical motion (with amplitude
$i_{\rm free}$ and vertical frequency $\nu_0$).
The free eccentricity $e_{\rm free}$ and inclination $i_{\rm free}$
are free parameters representing the amplitudes of the epicyclic and
vertical motion.
\cite{leu13} extended this theory to describe the orbit of a
test particle around an eccentric binary, to first order in the
eccentricity $e_{AB}$ of the binary. They found additional forced
oscillation terms with frequencies equal to $n_{AB}$ and $|k n_0 - (k
\pm 1) n_{AB}|$.
The solution for the cylindrical distance $R$ as a function of time is
\begin{eqnarray}
R(t)
&=& R_0\Bigg\{1 - e_{\rm free}\cos(\kappa_0t + \psi) - C_0\cos M_B  - \sum_{k=1}^{\infty}\Big[C_k^0\cos k(\phi_0 - M_B - \varpi_B)
\nonumber\\
&& + C_k^+\cos (k(\phi_0 - \varpi_B) - (k +1)M_B) 
 + C_k^-\cos (k(\phi_0 - \varpi_B) - (k -1)M_B)\Big]\Bigg\},  
\label{rt}
\end{eqnarray}
where $M_B = n_{AB}t + \varphi_{AB}$ and $\varpi_B$ are the mean
anomaly and longitude of periapse of the secondary of the
binary, $C_0$, $C_k^0$, and $C_k^{\pm}$ are amplitudes of the
forced oscillation terms, and $\varphi_{AB}$ is a constant.
The largest forced oscillation term is the forced eccentricity,
$e_{\rm forced} = C_1^-$, at frequency $n_0$.

The analytic expressions for the mean motion $n_0$, the epicyclic
frequency $\kappa_0$, the amplitudes $C_0$, $C_k^0$, and $C_k^{\pm}$
of the forced oscillation terms, and the  vertical frequency $\nu_0$
in Equations (21), (26), (28), (29), (30), and (36) of \cite{leu13},
respectively, are first order in $e_{AB}$ and in deviation from
circular motion.
In particular,
\begin{eqnarray}
n_0^2 
&=& {1 \over 2} \Bigg\{
    {m_A \over (m_A+m_B)} b_{1/2}^0(\alpha_{A}) +
    {m_B \over (m_A+m_B)} b_{1/2}^0(\alpha_{B}) \nonumber \\
& & \qquad + {m_A m_B \over (m_A + m_B)^2} \left(a_{AB} \over R_0\right)
             \left[Db_{1/2}^0(\alpha_{A}) + Db_{1/2}^0(\alpha_{B})
             \right]
\Bigg\} n_K^2 , \label{n0}
\end{eqnarray}
and
\begin{eqnarray}
\kappa_0^2
&=& {1 \over 2} \Bigg\{
    {m_A \over (m_A+m_B)} b_{1/2}^0(\alpha_{A}) +
    {m_B \over (m_A+m_B)} b_{1/2}^0(\alpha_{B})
\nonumber \\
& & \qquad - {m_A m_B \over (m_A + m_B)^2} \left(a_{AB} \over R_0\right)
             \left[Db_{1/2}^0(\alpha_{A}) + Db_{1/2}^0(\alpha_{B})
             \right]
\nonumber \\
& & \qquad - {m_A m_B \over (m_A + m_B)^2} \left(a_{AB} \over R_0\right)^2
             \Bigg[{m_B \over (m_A+m_B)} D^2b_{1/2}^0(\alpha_{A})
\nonumber \\
& & \qquad         + {m_A \over (m_A+m_B)} D^2b_{1/2}^0(\alpha_{B})\Bigg]
\Bigg\} n_K^2 ,
\label{kappa0}
\end{eqnarray}
where $a_{AB}$ is the semimajor axis of the orbit of the secondary of
mass $m_B$ relative to the primary of mass $m_A$, $\alpha_{A} =
a_A/R_0$, $\alpha_{B} = a_B/R_0$, $a_A = a_{AB} m_B/(m_A + m_B)$, $a_B
= a_{AB} m_A/(m_A + m_B)$, $b_s^k(\alpha)$ are the Laplace
coefficients, $D = d/d\alpha$, and $n_K = [G (m_A+m_B)/R_0^3]^{1/2}$
is the Keplerian mean motion at $R_0$.

How can we determine accurate values for these parameters, without the
approximations of the analytic expressions, and the free parameters
$e_{\rm free}$ and $i_{\rm free}$?
According to Equation ({\ref{rt}), if we apply FFT to $R(t)/R_0$ obtained
from numerical orbit integration, we expect that a peak with amplitude
$e_{\rm free}$ and power equal to the value of $e_{\rm free}^2/2$
would appear at the epicyclic frequency $\kappa_0$ in the power
spectrum.
In addition, we expect peaks at other frequencies, e.g.,
peaks with amplitudes $e_{\rm forced} = C_1^-$ at $n_0$, $C_1^0$ at
$|n_0-n_{AB}|$, $C_1^+$ at $|n_0-2n_{AB}|$, etc.
Among all the other peaks, $e_{\rm forced}$ is closest to $e_{\rm
  free}$ because of  the small difference between $n_0$ and
$\kappa_0$.
To separate these two peaks, we need a spectrum with frequency
resolution finer than $|n_0-\kappa_0|$.
FFT of $z(t)$ can also be used to determine $i_{\rm free}$ and
$\nu_0$, but we do not discuss the vertical motion below.

In order to test the feasibility of this method for different binary
systems, we apply FFT to three Kepler circumbinary planets and the
satellites of Pluto-Charon.
To obtain $R(t)$, we perform $N$-body simulations using the
Wisdom-Holman \citep{wis91} integrator in the SWIFT package 
\citep{lev94}, modified for integrations of systems with
comparable masses \citep{lee03}.
The output time interval and the integration time of the simulations
are chosen in order to generate a high resolution power spectrum.
The FFT code we used is taken from Section 13.4 of \cite{pre92}.
The frequency resolution of the power spectrum equals to
$\pi/(M\zeta)$, where $M$ is a power of 2 and depends on the length of
the data, and $\zeta$ is the output time interval of the data.
To obtain a high resolution spectrum, $M\zeta$ should be set to a
relatively large number.
However, $\zeta$ should not be larger than the orbital period of the
circumbinary object.
Otherwise, the output data may lose some information of the orbit. 
The amplitude that corresponds to a peak can be calculated by
$\sqrt{2\sum_kP_k}$, where $P_k$ are the powers at the points ($k= 1$,
2, or 3) that constitute a peak.

\section{RESULTS FROM FFT ANALYSIS}
\label{sec:result}

\subsection{Kepler Circumbinary Planets}

We apply FFT to three different circumbinary exoplanets:
Kepler-16~b, Kepler-34~b and Kepler-35~b.
These are the same systems studied by \cite{leu13}, and their orbital
parameters are listed in Table 1 of \cite{leu13}.
The systems are integrated for about 5000 years with an integration
time step $\delta t = 0.36525$ days and an output time step $\zeta =
40\delta t$.
Our aim is to separate the peaks at $n_0$ and $\kappa_0$.
Therefore, we need more than $10^5$ lines of output in order to obtain
a high enough resolution power spectrum.

Figures \ref{spec_kep16}, \ref{spec_kep34}, and \ref{spec_kep35} show the power
spectra of the three exoplanets. We observe that two peaks
corresponding to $e_{\rm free}$ and $e_{\rm forced}$ appear in the
spectra at frequencies $\kappa_0$ and $n_0$ close to the analytic
frequencies. Since the eccentricities of all three binary stars are
non-zero, we expect $e_{\rm forced}$ to exist in all three
spectra. For Kepler-16~b, its $e_{\rm free}$ and $e_{\rm forced}$ are
comparable with each other. However, $e_{\rm forced}$ is $\sim 100$
times smaller than $e_{\rm free}$ in Kepler-34~b and $\sim 10$ times
smaller than $e_{\rm free}$ in Kepler-35~b.
According to Equation (38) of \cite{leu13}, which is the lowest order
term in the expansion of the analytic expression for $e_{\rm forced} =
C_1^-$ in powers of $a_{AB}/R_0$,
\begin{equation}
e_{\rm forced}\approx\frac{5}{4}e_{AB}\frac{m_{A}-m_{B}}{m_{A}+m_{B}}\frac{a_{AB}}{R_0} .
\label{eforce}
\end{equation}
The reason for $e_{\rm forced}$ being small in the latter two systems
is that the stars within each system are comparable to each other
in mass, and $(m_A - m_B)/(m_A + m_B)$ is a small value.
Even though $e_{AB}$ of Kepler-34 and Kepler-35 are 0.52068 and 0.14224,
respectively, $e_{\rm forced}$ of Kepler-34~b and Kepler-35~b are
significantly smaller than that of Kepler-16~b (see Table
\ref{table1}). 

Table \ref{table1} shows a comparison between the values obtained from
the power spectra and those predicted by the first-order epicyclic
theory of \cite{leu13} for the three Kepler systems.
The frequencies $n_0$ and $\kappa_0$ are shown in units of the
Keplerian mean motion $n_K = [G(m_A + m_B)/R_0^3]^{1/2}$ at $R_0$.
The difference between the values of $e_{\rm forced}$
from the power spectra and the values from the epicyclic theory are
$\sim 0.08\%$, $\sim 12\%$ and $\sim 3\%$ for Kepler-16~b, Kepler-34~b
and Kepler-35~b, respectively. The large difference for Kepler-34~b is
caused by the nature of the epicyclic theory, which is only accurate to
first order in the binary eccentricity $e_{AB}$. If $e_{AB}$ is large,
as in Kepler-34, where $e_{AB} = 0.52068$, the analytic expressions of the epicyclic theory may not be accurate. For Kepler-16 and Kepler-35, $e_{AB}$ are only 0.16048 and 0.14224, respectively, which are more than 3 times smaller than $e_{AB}$ of Kepler-34.
We should emphasize that the parameter values from the power spectra
are more accurate than those from the analytic theory. This can be
illustrated by the periapse precession period of Kepler-34~b, which is
equal to $2\pi/|n_0 - \kappa_0|$. The analytic periapse precession
period is about 91 years, whereas the period calculated from the
spectrum is 62.4 years, which agrees with the numerical result shown
in the lower left panel of Figure 5 of \cite{leu13}.

\cite{leu13} estimated the free parameter $e_{\rm free}$ from the
variation of a transformed radius $R'$, where the forced
oscillation terms (including the forced eccentricity term) have been
evaluated using the analytic expressions of the first-order epicyclic
theory and subtracted from $R_0$.
These are the values shown in parenthesis in Table \ref{table1}.
The more accurate values determined from the power spectra are smaller
by $\sim 4$--$6$\%.

In Figures \ref{spec_kep16} to \ref{spec_kep35}, other than $e_{\rm free}$ and
$e_{\rm forced}$ at $\kappa_0$ and $n_0$, we also observe peaks at
frequencies which cannot be explained by the first-order epicyclic
theory. We tested and confirmed the reality of these peaks by changing
the output time steps and hence the resolution of the spectra, which
does not change the peaks. The identity of these peaks is still
unknown. We suspect that they may be due to higher order terms of $e_{AB}$ and forced oscillations, possibly with frequencies equal to some combinations of $n_0$ and $\kappa_0$. Further investigation is needed.   
There are also background fluctuations of the spectrum outside the
peaks.
We find that increasing the resolution of the spectrum decreases the
amplitude of the background fluctuations and allows the peaks to look
sharper.
Hence, the background fluctuations are likely to be artifacts of FFT.

\subsection{Satellites of Pluto-Charon}
\label{sec:pluto}

We also apply FFT to the satellites of Pluto-Charon.
We perform simulations with the initial state vectors of the best fit
of \cite{sho15}, with a time step $\delta t$ of $3000\,$s.
The output time step $\zeta$ is chosen to be 30 times $\delta
t$ and the system is integrated for $\sim 1200$ days in order to
generate high resolution spectra that can separate $n_0$ and
$\kappa_0$.

Figures \ref{spec_styx} and \ref{spec_nix} show the power spectra of
Styx and Nix, respectively.
We observe two peaks in the power spectra near the analytic values of
$n_0$ and $\kappa_0$.
The peak near $n_0$ is not expected, because the orbit of Pluto-Charon
is circular ($e_{AB} = 0$) and $e_{\rm forced}$ should be zero,
according to the first-order epicyclic theory (see Equation (30) of
\cite{leu13} and Equation (\ref{eforce}) above).
This peak is likely due to higher order terms, but since we are not
certain of its origin, we do not use the position of this peak to
determine $n_0$ from the numerical integration.
Instead, we use the cumulative increase in $\phi$, as in \cite{lee06}.
The values of $\kappa_0$ and $e_{\rm free}$ are determined from the
power spectrum.

Table \ref{table2} shows the numerical values of $n_0/n_K$,
$\kappa_0/n_K$, and $e_{\rm free}$ for all four satellites.
They are compared to the analytic values of $n_0/n_K$ and
$\kappa_0/n_K$ from the first-order epicyclic theory and the best-fit
values from Table 1 of \cite{sho15}.
The latter were obtained from an $8$-parameter fit of a precessing
ellipse to the entire set of observational data, and we converted the
apsidal precession rate $\dot\varpi$ to $\kappa_0$ using $\kappa_0 =
n_0 - {\dot\varpi}$.
For Nix, Kerberos, and Hydra, the values of $n_0/n_K$ and
$\kappa_0/n_K$ by all three methods are in good agreement with each
other, except that the analytic value of $\kappa_0/n_K$ is slightly
smaller for Nix, and the values of $e_{\rm free}$ from the power
spectrum are smaller than the eccentricity from the precessing ellipse
fit by $\sim 3$--$9\%$.
However, for Styx, the satellite closest to Pluto-Charon, the value of
$\kappa_0/n_K$ from the power spectrum is significantly larger than
the analytic and best-fit values (see also Figure \ref{spec_styx}),
and the value of $e_{\rm free}$ is more than five times smaller than
the best-fit value.

In addition to the $8$-parameter fit shown in their Table 1, where the
nodal precession rate is derived from $n$ and $\dot\varpi$,
\cite{sho15} also explored fits to the entire data set with $6$--$9$
parameters that make different assumptions about how the parameters
are coupled, and the resulting orbiting elements are listed in their
Extended Table 1.
Interestingly, for Nix, Kerberos, and Hydra, all of the fits give
similar results for $e$ and $\dot\varpi$ (and hence $\kappa_0$), but
for Styx, the fits with $6$ and $7$ parameters give results that are
different from those with $8$ and $9$ parameters.
For $6$ and $7$ parameters, $e \approx 0.0011$, which is close to the
value we find for $e_{\rm free}$ from the power spectrum, and
${\dot\varpi} = 0.37688^\circ\,{\rm day}^{-1}$ or $\kappa_0/n_K
\approx 0.9873$, which is larger than $\kappa_0/n_K \approx 0.9800$
for the $8$- and $9$-parameter fits but still smaller than
$\kappa_0/n_K = 0.99969$ from the power spectrum.

To get a better understanding of this disagreement in the numerical
and analytic values of $\kappa_0$ for Styx, we pick some test
particles from our study of the early in-situ formation scenario of
the small satellites \citep{woo18} and apply FFT to their final orbits
after the tidal evolution of Pluto-Charon.
They are taken from the constant $\Delta t$ simulations shown in
Figure 10 of \cite{woo18}, with the time lag of the tidal bulge of
Pluto $\Delta t = 600\,$s, the ratio of the rates of tidal dissipation
in Charon and Pluto $A = 40$, and the initial $e_{AB} = 0$ (hence
$e_{AB} = 0$ from beginning to end), and they have small $e_{\rm
  free}$ ($\la 10^{-4}$). 
Their $\kappa_0/n_K$ from the power spectrum and $n_0/n_K$ from the
cumulative increase in $\phi$ are shown as black dots in Figure
\ref{kappa_n}.
The analytic expression from the first-order epicyclic theory
(Equations (\ref{n0}) and (\ref{kappa0})) are shown as black lines.
As we can see, their $n_0$ and $\kappa_0$ follow the same trends as the
actual satellites (magenta alphabets in Figure \ref{kappa_n}), with
$n_0$ in good agreement with the analytic
expression and $\kappa_0$ showing increasing deviation from the
analytic expression as $R_0$ decreases.
We shall seek an explanation for this deviation for $\kappa_0$ in
Section \ref{sec:theory}.

\subsection{Forced Oscillation Terms}

Beside $e_{\rm free}$ and $e_{\rm forced}$, we also search for the
peaks of the forced oscillation terms, which are located in higher
frequency regions of the power spectrum, compared to $n_0$ and
$\kappa_0$. Figure \ref{spec_nix_2} shows the forced oscillation terms of
Nix, and Figure \ref{spec_kep16_2} shows the forced oscillation terms of
Kepler-16~b. More identifiable peaks are found in Figure \ref{spec_kep16_2}
than in Figure \ref{spec_nix_2}, which can be explained by the epicyclic theory.
From Equations (28) and (30) of \cite{leu13}, $C_0$ and $C_k^\pm$ are
proportional to $e_{AB}$ and exist only when $e_{AB}$ is non-zero.
Hence, we cannot find peaks of $C_0$, $C_2^-$ and $C_3^-$ in Figure
\ref{spec_nix_2} but they exist in Figure \ref{spec_kep16_2}. We do not search for $C_k^+$ for Kepler-16~b, because their values are much smaller than the values of $C_k^0$ and $C_k^-$, for $k$ = 1 to 3 (see Table 2 of \citealt{leu13}). 
There are many smaller peaks in Figures \ref{spec_nix_2} and \ref{spec_kep16_2} at
frequencies other than those associated with $C_k^0$, $C_0$ and
$C_k^\pm$, whose identities are unknown but again could be due to
higher order terms of $e_{AB}$ and forced oscillations.

We compare the amplitudes and frequencies of the forced
oscillation terms from the power spectrum to their predicted values
from the first-order epicyclic theory.
The peaks are found at frequencies close to their analytic
predictions.
Table \ref{table3} shows the comparison between the amplitudes from the
power spectrum and the analytic values from \cite{leu13} for some of
the forced oscillation terms of Kepler-16~b. We find that the
amplitudes from the power spectrum are slightly lower than the
analytic values. Similar to $e_{\rm forced}$, the slight difference is
expected to be due to the inaccurate prediction from the first-order
epicyclic theory of \cite{leu13}.

\subsection{Uncertainties}

The FFT results presented in the above subsections for each
circumbinary object are obtained from a numerical orbit integration of
the best fit using a specific time step.
The uncertainties due to the numerical errors of the orbit integration
can be estimated by varying the time step.
In particular, we find that the uncertainty in $e_{\rm free}$ is less
than $0.4\%$ for Styx and less than $0.05\%$ for the other satellites
of Pluto-Charon.
The uncertainties due to the uncertainties in the measurements can be
determined by applying the FFT method to numerical orbit integrations
of the distribution of fits from a Markov chain Monte Carlo or
bootstrap analysis.
Such an analysis is beyond the scope of the present paper.

\section{MODIFICATION OF EPICYCLIC FREQUENCY BY 3:1
  MEAN-MOTION RESONANCE}
\label{sec:theory}

How can we explain the increasing deviation of $\kappa_0$ from the
lowest order epicyclic theory at decreasing distance from the center
of mass of Pluto-Charon seen in Figure \ref{kappa_n}?
For motion around an oblate planet with $J_2$ and $J_4$, \cite{bor94}
have extended the epicyclic theory to third order and found
corrections to $n_0$, $\kappa_0$, and $\nu_0$ proportional to
$e_{\rm free}^2$ and $i_{\rm free}^2$.
Similar corrections from the axisymmetric components of the potential
of a binary are negligible for the satellites and test particles shown
in Figure \ref{kappa_n}, which have small $e_{\rm free}$ and
$i_{\rm free}$.
Generalizing this approach to account for the time varying
non-axisymmetric components of the potential of a binary is difficult,
with many forced oscillation terms to keep track of just to go from
first order to second order.

Instead, we follow the Hamiltionian approach of \cite{lit08}.
For a particle of mass $m$ orbiting a binary with primary mass $m_A$
and secondary mass $m_B$, the Hamiltonian in Jacobi coordinates is
\begin{equation}
{\cal H} = {\cal H}_{K, AB} + {\cal H}_{K} + {\cal H}_{\rm pert} ,
\end{equation}
where
\begin{equation}
{\cal H}_{K, AB} = - {G m_A m_B \over 2 a_{AB}}
\end{equation}
represents the Keplerian motion of the binary with osculating
semimajor axis $a_{AB}$,
\begin{equation}
{\cal H}_{K} = - {G (m_A + m_B) m \over 2 a}
\end{equation}
represents the Keplerian motion of the particle relative to the
center of mass of the binary with osculating semimajor axis $a$,
and
\begin{equation}
{\cal H}_{\rm pert} = - G m m_A \left({1 \over {|\vecr - \vecr_A|}}
                                       - {1 \over |\vecr|}\right) 
                       - G m m_B \left({1 \over {|\vecr - \vecr_B|}}
                       - {1 \over |\vecr|}\right)
\label{Hpert}
\end{equation}
represents the perturbations to the Keplerian motions \citep{lee03}.
We shall consider the limit $m \ll m_A + m_B$, where the binary orbit
is fixed.

In Equation (\ref{Hpert}), $\vecr$, $\vecr_A = - m_B \vecr_{AB}/(m_A +
m_B)$, and $\vecr_B = m_A \vecr_{AB}/(m_A + m_B)$ are the positions of
the particle, primary, and secondary from the center of mass of the
binary, respectively, where $\vecr_{AB}$ is the position of the
secondary from the primary.
For the terms $1/|\vecr - \vecr_A|$ and  $1/|\vecr - \vecr_B|$, we can
use the usual expansion of the direct part of the disturbing function
$R_D = a/|\vecr - \vecr'|$ for an outer body with $r > r'$ into
various terms given in, e.g., Chapter 6.5 of \cite{mur99}.
For the term $1/|\vecr|$, we can use the same expansion and take the
limit $a'/a \to 0$.
If we assume that $e_{AB} = 0$ and that the orbits are coplanar,
we then find that
\begin{equation}
{\cal H}_{\rm pert} = {\cal H}_{\rm sec} + {\cal H}_{\rm 3:1} + \cdots,
\end{equation}
where
\begin{eqnarray}
{\cal H}_{\rm sec} &=&
  - {G m m_A \over a} \left[{1\over 2} b_{1/2}^{0}(\alpha) +
    {1 \over 8} e^2 (2 \alpha D + \alpha^2 D^2)
    b_{1/2}^{0}(\alpha) - 1\right]_{\alpha_A'}
\nonumber\\
&&- {G m m_B \over a} \left[{1\over 2} b_{1/2}^{0}(\alpha) +
    {1 \over 8} e^2 (2 \alpha D + \alpha^2 D^2)
    b_{1/2}^{0}(\alpha) - 1\right]_{\alpha_B'}
\end{eqnarray}
is the secular term, and
\begin{eqnarray}
{\cal H}_{\rm 3:1} &=&
  + {G m m_A \over a} \left[{1 \over 8} e^2 (17 + 10 \alpha D + \alpha^2 D^2)
    b_{1/2}^{1}(\alpha)\right]_{\alpha_A'} \cos(3\lambda - \lambda_{AB} - 2 \varpi)
\nonumber\\
&&- {G m m_B \over a} \left[{1 \over 8} e^2 (17 + 10 \alpha D + \alpha^2 D^2)
    b_{1/2}^{1}(\alpha)\right]_{\alpha_B'} \cos(3\lambda - \lambda_{AB} - 2 \varpi) 
\end{eqnarray}
is the 3:1 mean-motion resonance term.
The expressions in the square brackets are evaluated at $\alpha =
\alpha_A' = a_{A}/a$ and $\alpha = \alpha_B' = a_{B}/a$, $\lambda =
n_0 t + \psi_0$ and $\lambda_{AB} = n_{AB}t + \varphi_{AB} + \varpi_B$.
Unlike \cite{lit08}, we do not expand the above expressions to leading
order in $m_B/(m_A + m_B)$.

From the equation of motion for $\lambda$,
\begin{equation}
n_0 = {d\lambda \over dt} = {2 \over m} \sqrt{a \over G (m_A+m_B)}
                {\partial {\cal H} \over \partial a} 
\end{equation}
with the Keplerian and secular terms: ${\cal H} = {\cal H}_K +
{\cal H}_{\rm sec}$, we find that
\begin{eqnarray}
n_0
&=& \Bigg\{
    -1 + {m_A \over (m_A+m_B)} b_{1/2}^0(\alpha_{A}') +
    {m_B \over (m_A+m_B)} b_{1/2}^0(\alpha_{B}') \nonumber \\
& & \quad + {m_A m_B \over (m_A + m_B)^2} \left(a_{AB} \over a\right)
             \left[Db_{1/2}^0(\alpha_{A}') + Db_{1/2}^0(\alpha_{B}')
             \right] \nonumber \\
& & \quad + {1 \over 8}{m_A m_B \over (m_A + m_B)^2} \left(a_{AB} \over a\right)
             e^2 \Big[4 Db_{1/2}^0(\alpha_{A}') + 5 \alpha_A' D^2
               b_{1/2}^0(\alpha_{A}') + \alpha_A'^3 D^2
               b_{1/2}^0(\alpha_{A}') \nonumber \\
& & \qquad     + 4 Db_{1/2}^0(\alpha_{B}') + 5 \alpha_B' D^2
               b_{1/2}^0(\alpha_{B}') + \alpha_B'^3 D^2
               b_{1/2}^0(\alpha_{B}')
             \Big]\Bigg\} n_a ,
\label{n0sec}
\end{eqnarray}
where $n_a = [G (m_A + m_B)/a^3]^{1/2}$.
The last term in Equation (\ref{n0sec}) is proportional to $e^2$,
which is similar to the correction found by \cite{bor94} and can be
neglected for the satellites of Pluto-Charon with small $e$.
As \cite{gre81} pointed out, if we consider only the axisymmetric
components of the potential, a test particle on a circular orbit at
$R_0$ is always at periapse in the osculating elements, i.e., $R_0 =
a_{\rm osc} (1 - e_{\rm osc})$.
We can then use the angular momentum of a circular orbit at $R_0$,
$m R_0^2 n_0 = m \left[G (m_A + m_B) a_{\rm osc} (1 - e_{\rm osc}^2)\right]^{1/2}$,
to derive that $e_{\rm osc} = n_0^2/n_K^2 - 1$ and $a_{\rm osc} =
R_0/(2- n_0^2/n_K^2)$, where $n_K = [G (m_A + m_B)/R_0^3]^{1/2}$ as before.
Thus the appropriate expression to use for $a$ in Equation
(\ref{n0sec}) is $a = R_0/(2- n_0^2/n_K^2)$.
This $n_0$ is plotted as the upper red curve in Figure \ref{kappa_n},
which is in good agreement with the upper black curve showing Equation
(\ref{n0}) from the first-order epicyclic theory.
It can also be shown analytically that Equations (\ref{n0}) and
(\ref{n0sec}) (without the term proportional to $e^2$ in the latter)
are equivalent to each other to the lowest order in the deviation of
$n_0$ from $n_K$.

For the complex eccentricity $z = e \exp(i\varpi)$, the equation of
motion (to the lowest order in $e$) is
\begin{equation}
{dz \over dt} = - {2i \over m \sqrt{G (m_A+m_B) a}}
                {\partial {\cal H} \over \partial z^\ast}
\end{equation}
If we consider only the secular term ${\cal H}_{\rm sec}$,
\begin{equation}
{dz \over dt} = i \eta z ,
\end{equation}
where
\begin{eqnarray}
\eta
&=& {1 \over 4} {m_A m_B \over (m_A + m_B)^2}
    \left(a_{AB} \over a\right) \Bigg\{2 D b_{1/2}^0(\alpha_A')
    + 2 D b_{1/2}^0(\alpha_B') \nonumber \\
& & \quad + \left(a_{AB} \over a\right)\left[{m_A \over (m_A + m_B)} D^2
      b_{1/2}^0(\alpha_B') + {m_B \over (m_A + m_B)} D^2
      b_{1/2}^0(\alpha_A')\right]\Bigg\} n_a.
\end{eqnarray}
The solution is
\begin{equation}
z = e_{\rm free} \exp(i\eta t) .
\end{equation}
So the secular periapse precession rate is ${\dot \varpi} = \eta$ and
the epicyclic frequency is $\kappa_0 = n_0 - {\dot\varpi} = n_0 - \eta$.
This $\kappa_0$ is plotted as the lower red curve in Figure \ref{kappa_n},
using Equation (\ref{n0sec}) for $n_0$ and $a = R_0/(2- n_0^2/n_K^2)$, and
it is in good agreement with the lower black curve showing Equation
(\ref{kappa0}) from the first-order epicyclic theory.
Again it can be shown analytically that $\kappa_0$ from the first-order
epicyclic theory and the secular Hamiltonian are equivalent to each
other to the lowest order in the deviation of $\kappa_0$ from $n_K$.

\cite{lit08} showed that the 3:1 mean-motion resonance term
${\cal H}_{\rm 3:1}$ can affect the rate of tidal damping even far
from the nominal position of the resonance.
Their derivation indicates that the 3:1 term can also affect the
periapse precession rate.
If we include the 3:1 resonance term ${\cal H}_{\rm 3:1}$ in the
equation of motion for $z$,
\begin{equation}
{dz \over dt} = i \eta z + i \nu z^\ast \exp\{i[(3 n_0 - n_{AB}) t +
  \Delta\varphi]\} ,
\end{equation}
where $\Delta\varphi = 3 \psi_0 - \varphi_{AB} - \varpi_B$ and
\begin{eqnarray}
\nu
&=& {1 \over 4} \Bigg\{{m_B \over (m_A + m_B)} \left[17
  b_{1/2}^1(\alpha_B') + 10 \alpha_B' D b_{1/2}^1(\alpha_B') +
  \alpha_B'^2 D^2 b_{1/2}^1(\alpha_B') \right] \nonumber \\
& & \quad - {m_A \over (m_A + m_B)} \left[17
  b_{1/2}^1(\alpha_A') + 10 \alpha_A' D b_{1/2}^1(\alpha_A') +
  \alpha_A'^2 D^2 b_{1/2}^1(\alpha_A') \right] \Bigg\} n_a .
\end{eqnarray}
This equation has the solution $z = k \exp[i (\eta + b) t] + k (b/\nu)
\exp[i (3 n - n_{AB} - \eta - b) t + \Delta\varphi]$, where $b$
satisfies the quadratic equation
\begin{equation}
b^2 - (3 n - n_{AB} - 2\eta) b + \nu^2 = 0 ,
\end{equation}
which has the roots
\begin{equation}
b = {(3 n - n_{AB} - 2 \eta) \pm \sqrt{(3 n - n_{AB} - 2 \eta)^2 - 4
    \nu^2}\over 2} .
\end{equation}
The relevant root is the one with the positive sign.
Hence
\begin{equation}
z = e_{\rm free} \exp[i (\eta + b) t] + e_{\rm free} (b/\nu) \exp\{i
[(3 n - n_{AB} - b - \eta) t + \Delta\varphi]\} ,
\end{equation}
the periapse precession rate ${\dot \varpi} = \eta + b$, and the
epicyclic frequency $\kappa_0 = n_0 - {\dot\varpi} = n_0 - \eta - b$.
This modified epicyclic frequency is plotted as the green curve in
Figure \ref{kappa_n}.
The effect of the 3:1 resonance is non-trivial even at the position of
Nix near the nominal location of the 4:1 resonance, and it increases
with decreasing distance to the nominal location of the 3:1 resonance.
It explains a significant fraction of the discrepancy between the
epicyclic frequencies from the power spectrum for the satellites and
test particles and the epicyclic frequencies from the first-order
epicyclic theory and the secular theory.
The remaining discrepancy is likely due to other terms in the
perturbation Hamiltonian that we have not considered.

\section{SUMMARY AND DISCUSSION}
\label{sec:con}

In this paper, we have shown that the power spectrum generated by FFT
of $R(t)/R_0$ from numerical orbit integration is a powerful tool for
extracting the orbital parameters (such as $e_{\rm free}$,
$e_{\rm forced}$, $\kappa_0$, $n_0$, etc.) of a circumbinary object.
We applied this method to three Kepler circumbinary planets and the
satellites of Pluto-Charon.
The parameters extracted by the FFT power spectrum method are more
accurate than the analytic values from the first-order epicyclic
theory.
This can be important in applications to circumbinary exoplanets where
the binary eccentricity is large, as illustrated by the example of
Kepler-34~b.
This method is also able to determine accurately $e_{\rm free}$, which
is a free parameter in the epicyclic theory.

The FFT method was used by \cite{woo18} to measure $e_{\rm free}$ of
test particles at the end of simulations of the early in-situ
formation scenario of the small satellites of Pluto-Charon, which were
compared to the eccentricities of the small satellites reported by
\cite{sho15}.
In this paper, we have re-determined the orbital parameters of the
small satellites by using the FFT method and found significant
discrepancies in the values of $\kappa_0$ and $e_{\rm free}$ for Styx.
The discrepancy in $\kappa_0$ means that the numerical value of the
periapse precession rate ${\dot\varpi} = n_0 - \kappa_0$ is only
$41\%$ of the analytic value from the first-order epicyclic theory,
while the numerical value of $e_{\rm free}$ is more than five times
smaller than the eccentricity reported by \cite{sho15}.
By applying the FFT method to determine $\kappa_0$ for some test
particles in the simulations of \cite{woo18}, we showed that there is
a trend of increasing deviation of $\kappa_0$ from the analytic value
of the first-order epicyclic theory as the distance from the center of
mass of Pluto-Charon decreases.
We have developed an analytic theory based on the disturbing function
approach, which shows that this deviation in $\kappa_0$ is due to the
modification of the secular precession frequency by the 3:1
mean-motion resonance.

The lower value of $e_{\rm free}$ for Styx also has implications for
the origin and evolution of the satellites of Pluto-Charon.
According to Table 1 of \cite{sho15}, the eccentricity of Styx is
similar to that of Hydra, while the eccentricity increases smoothly
from Nix to Kerberos to Hydra (see also Table \ref{table2}).
\cite{woo18} found that these eccentricities are similar to those of
test particles that form in situ and survive to the end of the tidal
evolution of Pluto-Charon in a model with constant tidal dissipation
function $Q$, $Q_P = 100$ for Pluto, the relative rate of tidal
dissipation in Charon and Pluto $A = 2.5$, and initial orbital
eccentricity of Charon $e_{AB} = 0.2$ (although this model does not
explain the near resonance locations of the satellites).
The revised, much lower $e_{\rm free}$ for Styx is no longer
consistent with this model.
It is unclear whether changes in $Q_P$, $A$ and/or initial $e_{AB}$
could result in eccentricities after tidal evolution that are
consistent with the revised $e_{\rm free}$ of the satellites.
Alternatively, it is interesting to note that the revised
$e_{\rm free}$ increases smoothly from Styx to Hydra (see Table
\ref{table2}), which could be consistent with the cumulative effects
of passing Kuiper belt objects on the orbits of the satellites
\citep{col08}.
Further investigations into the viability of these scenarios are
needed.

\acknowledgments
We thank Mark Showalter for providing the initial state vectors of
Pluto's satellites from Showalter \& Hamilton and Yoram Lithwick and
Yanqin Wu for informative discussions.
This work was supported by a postgraduate studentship at the
University of Hong Kong (M.Y.W.) and Hong Kong RGC grant HKU 7030/11P
(M.Y.W. and M.H.L.).

\clearpage

\begin{deluxetable}{lrr}
\tablewidth{0pt}
\tablecaption{Comparison of Parameter Values from Power Spectrum and
  Epicyclic Theory for Three Kepler Circumbinary Planets
\label{table1}}
\tablehead{
\colhead{Parameter} & \colhead{Numerical} & \colhead{Analytic}
}
\startdata
\multicolumn{3}{c}{Kepler-16 b}\\
\noalign{\vskip .7ex}
\cline{1-3}
\noalign{\vskip .7ex}
$n_K$ (${\rm day}^{-1}$) & \multicolumn{2}{c}{0.02760}\\
$n_0/n_K$ & 1.00858 & 1.00702\\
$\kappa_0/n_K$ & 0.99574 & 0.99224\\
$e_{\rm free}$ & 0.0282 & (0.030) \\
$e_{\rm forced}$ & 0.0358 & 0.0358 \\
\noalign{\vskip .7ex}
\cline{1-3}
\noalign{\vskip .7ex}
\multicolumn{3}{c}{Kepler-34 b}\\
\noalign{\vskip .7ex}
\cline{1-3}
\noalign{\vskip .7ex}
$n_K$ (${\rm day}^{-1}$) & \multicolumn{2}{c}{0.02204}\\
$n_0/n_K$ & 1.00696 & 1.00423\\
$\kappa_0/n_K$ & 0.99445 & 0.99567\\
$e_{\rm free}$ & 0.193 & (0.204) \\
$e_{\rm forced}$ & 0.00164 & 0.00186 \\
\noalign{\vskip .7ex}
\cline{1-3}
\noalign{\vskip .7ex}
\multicolumn{3}{c}{Kepler-35 b}\\
\noalign{\vskip .7ex}
\cline{1-3}
\noalign{\vskip .7ex}
$n_K$ (${\rm day}^{-1}$) & \multicolumn{2}{c}{0.04897}\\
$n_0/n_K$ & 1.00825 & 1.00838\\
$\kappa_0/n_K$ & 0.99190 & 0.99119\\
$e_{\rm free}$ & 0.0364 & (0.038) \\
$e_{\rm forced}$ & 0.00242 & 0.00249 \\
\enddata
\tablecomments{The values of $e_{\rm free}$ in parenthesis were
  estimated from the variation of a transformed radius $R'$ by
  \cite{leu13} (see text for details).}
\end{deluxetable}

\begin{deluxetable}{lrrr}
\tablewidth{0pt}
\tablecaption{Comparison of Parameter Values from Power Spectrum,
  Epicyclic Theory, and Fit for Pluto-Charon Satellites
\label{table2}}
\tablehead{
\colhead{Parameter} & \colhead{Numerical} & \colhead{Analytic} & \colhead{Fit\tablenotemark{a}}
}
\startdata
\multicolumn{4}{c}{Styx}\\
\noalign{\vskip .7ex}
\cline{1-4}
\noalign{\vskip .7ex}
$n_K$ (${\rm day}^{-1}$) & \multicolumn{3}{c}{0.30898}\\
$n_0/n_K$ & 1.00849 & 1.00903 & 1.00863\\
$\kappa_0/n_K$ & 0.99969 & 0.98770 & 0.98004\\
$e_{\rm free}$ & 0.00110 & \nodata & 0.00579\\
\noalign{\vskip .7ex}
\cline{1-4}
\noalign{\vskip .7ex}
\multicolumn{4}{c}{Nix}\\
\noalign{\vskip .7ex}
\cline{1-4}
\noalign{\vskip .7ex}
$n_K$ (${\rm day}^{-1}$) & \multicolumn{3}{c}{0.25117}\\
$n_0/n_K$ & 1.00650 & 1.00658 & 1.00649\\
$\kappa_0/n_K$ & 0.99306 & 0.99171 & 0.99377\\
$e_{\rm free}$ & 0.00187 & \nodata & 0.00204 \\
\noalign{\vskip .7ex}
\cline{1-4}
\noalign{\vskip .7ex}
\multicolumn{4}{c}{Kerberos}\\
\noalign{\vskip .7ex}
\cline{1-4}
\noalign{\vskip .7ex}
$n_K$ (${\rm day}^{-1}$) & \multicolumn{3}{c}{0.19445}\\
$n_0/n_K$ & 1.00438 & 1.00452 & 1.00451\\
$\kappa_0/n_K$ & 0.99494 & 0.99468 & 0.99419\\
$e_{\rm free}$ & 0.00320 & \nodata & 0.00328\\
\noalign{\vskip .7ex}
\cline{1-4}
\noalign{\vskip .7ex}
\multicolumn{4}{c}{Hydra}\\
\noalign{\vskip .7ex}
\cline{1-4}
\noalign{\vskip .7ex}
$n_K$ (${\rm day}^{-1}$) & \multicolumn{3}{c}{0.16389}\\
$n_0/n_K$ & 1.00373 & 1.00354 & 1.00357\\
$\kappa_0/n_K$ & 0.99627 & 0.99598 & 0.99612\\
$e_{\rm free}$ & 0.00551 & \nodata & 0.00586\\
\enddata
\tablenotetext{a}{Best-fit values from Table 1 of \cite{sho15},
  with $\kappa_0 = n_0 - {\dot\varpi}$.}
\end{deluxetable}

\begin{deluxetable}{lrr}
\tablewidth{0pt}
\tablecaption{Comparison of Values of Some Forced Oscillation Terms from Power Spectrum and Epicyclic Theory for Kepler-16 b
\label{table3}}
\tablehead{
\colhead{Parameter} & \colhead{Numerical} & \colhead{Analytic}
}
\startdata
$|C_1^0|$ & 0.000248 & 0.000282 \\
$|C_2^0|$ & 0.000553 & 0.000589 \\
$|C_0|$   & 0.000135 & 0.000159\\
$|C_2^-|$ & 0.002358 & 0.002438\\
$|C_3^-|$ & 0.000085 & 0.000110\\
\enddata
\end{deluxetable}

\clearpage

\begin{figure}
\begin{center}
\includegraphics[angle=270,width=0.7\textwidth]{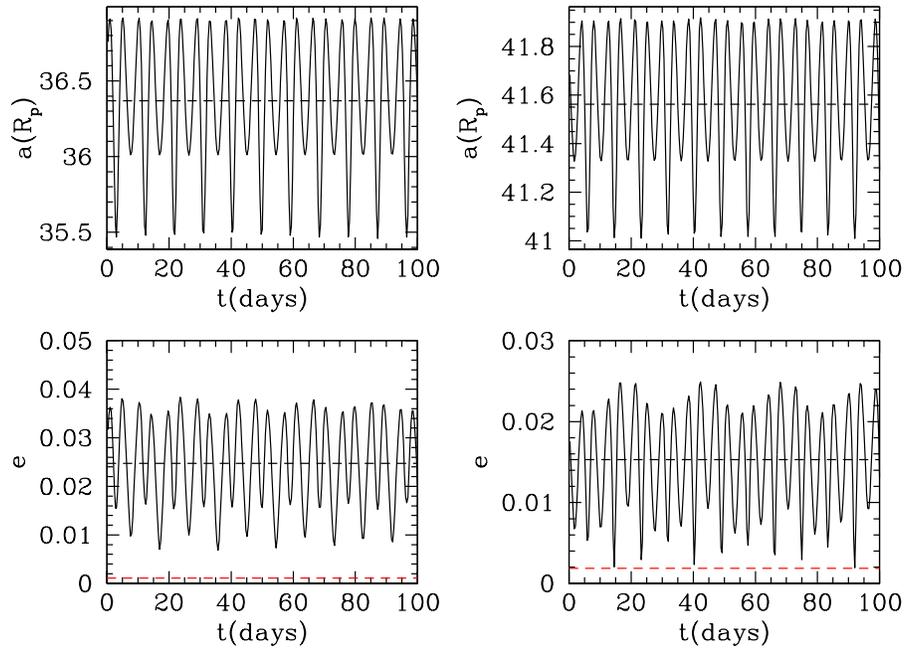}
\caption{Evolution of the osculating semimajor axis $a_{\rm osc}$ in
units of Pluto radius $R_P$ (upper panels) and eccentricity
$e_{\rm osc}$ (lower panels) of the Pluto-Charon satellites Styx (left
panels) and Nix (right panels).
The black dashed lines indicate the mean values of $a_{\rm osc}$ and
$e_{\rm osc}$, and the red dashed lines indicate the values of the
free eccentricity $e_{\rm free}$.
\label{aeosc}
}
\end{center}
\end{figure}


\begin{figure}
\begin{center}
\includegraphics[angle=270,width=0.75\textwidth]{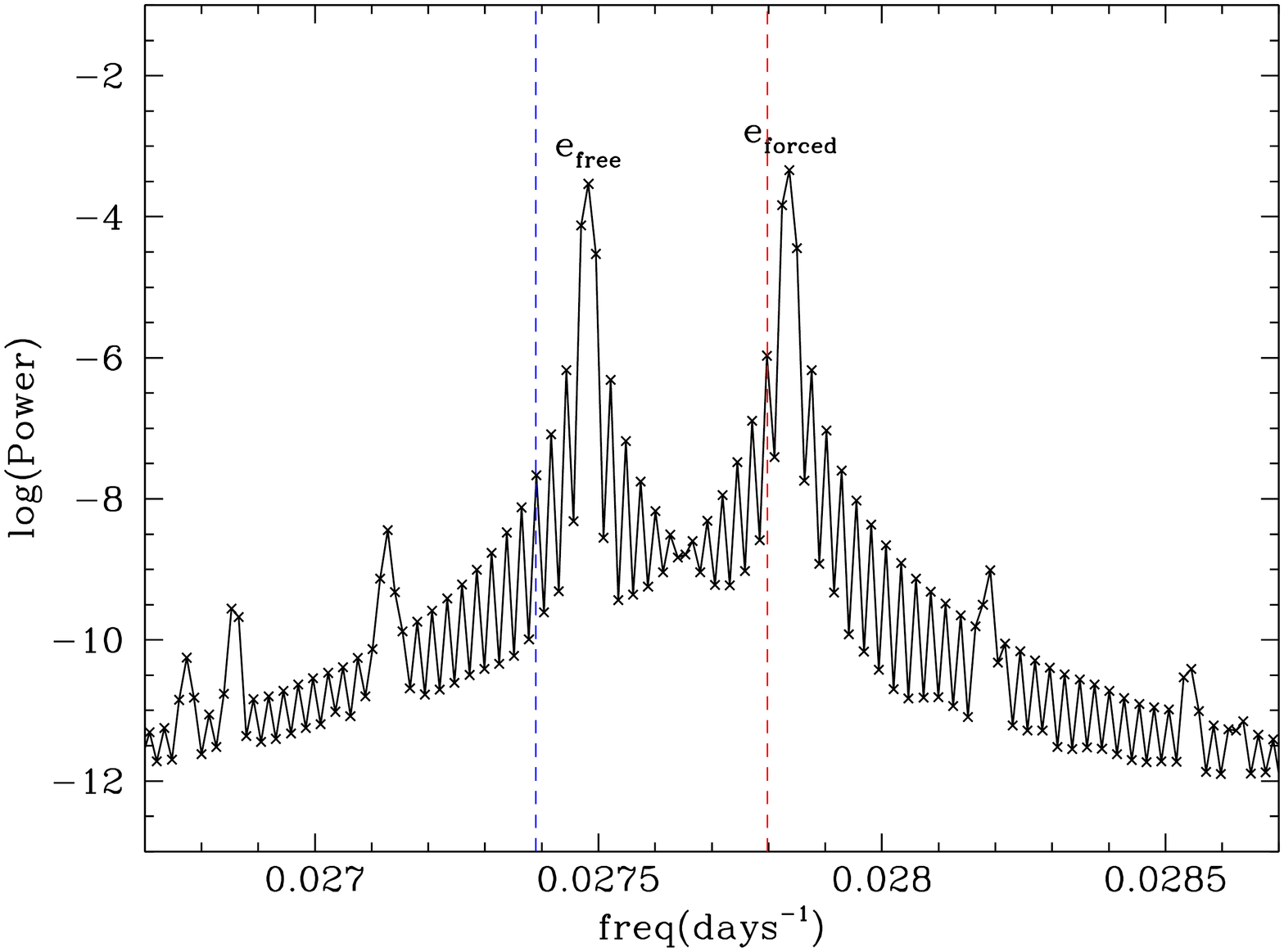}
\caption{Power spectrum obtained from FFT of $R(t)/R_0$ for Kepler-16 b.
See Table \ref{table1} for the parameter values. The vertical blue and
red dashed lines indicate the analytic values of $\kappa_0$ and $n_0$,
respectively, from the first-order epicyclic theory.
\label{spec_kep16}
}
\end{center}
\end{figure}


\begin{figure}
\begin{center}
\includegraphics[angle=270,width=0.75\textwidth]{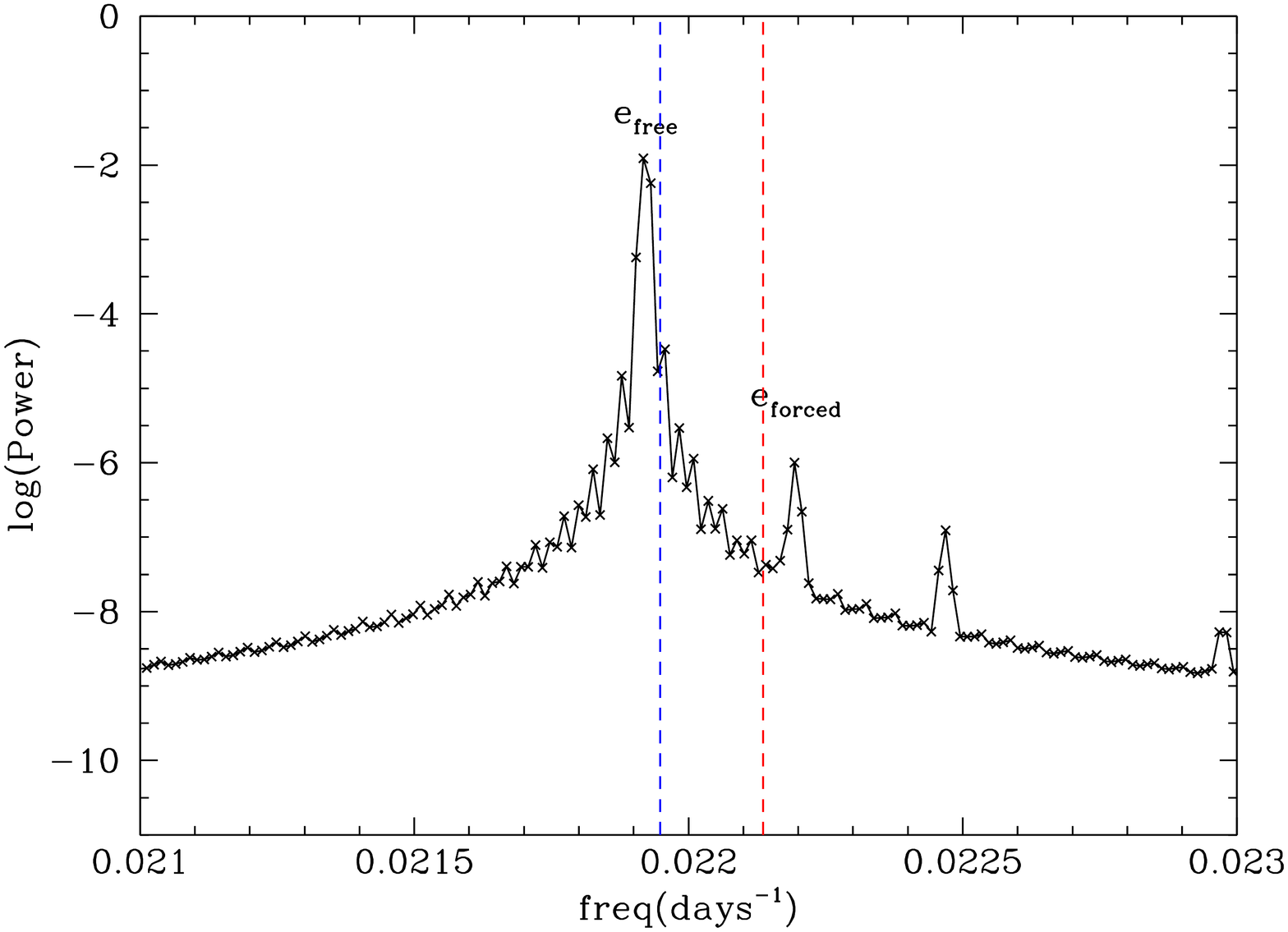}
\caption{Same as Figure \ref{spec_kep16}, but for Kepler-34 b. 
\label{spec_kep34}
}
\end{center}
\end{figure}


\begin{figure}
\begin{center}
\includegraphics[angle=270,width=0.75\textwidth]{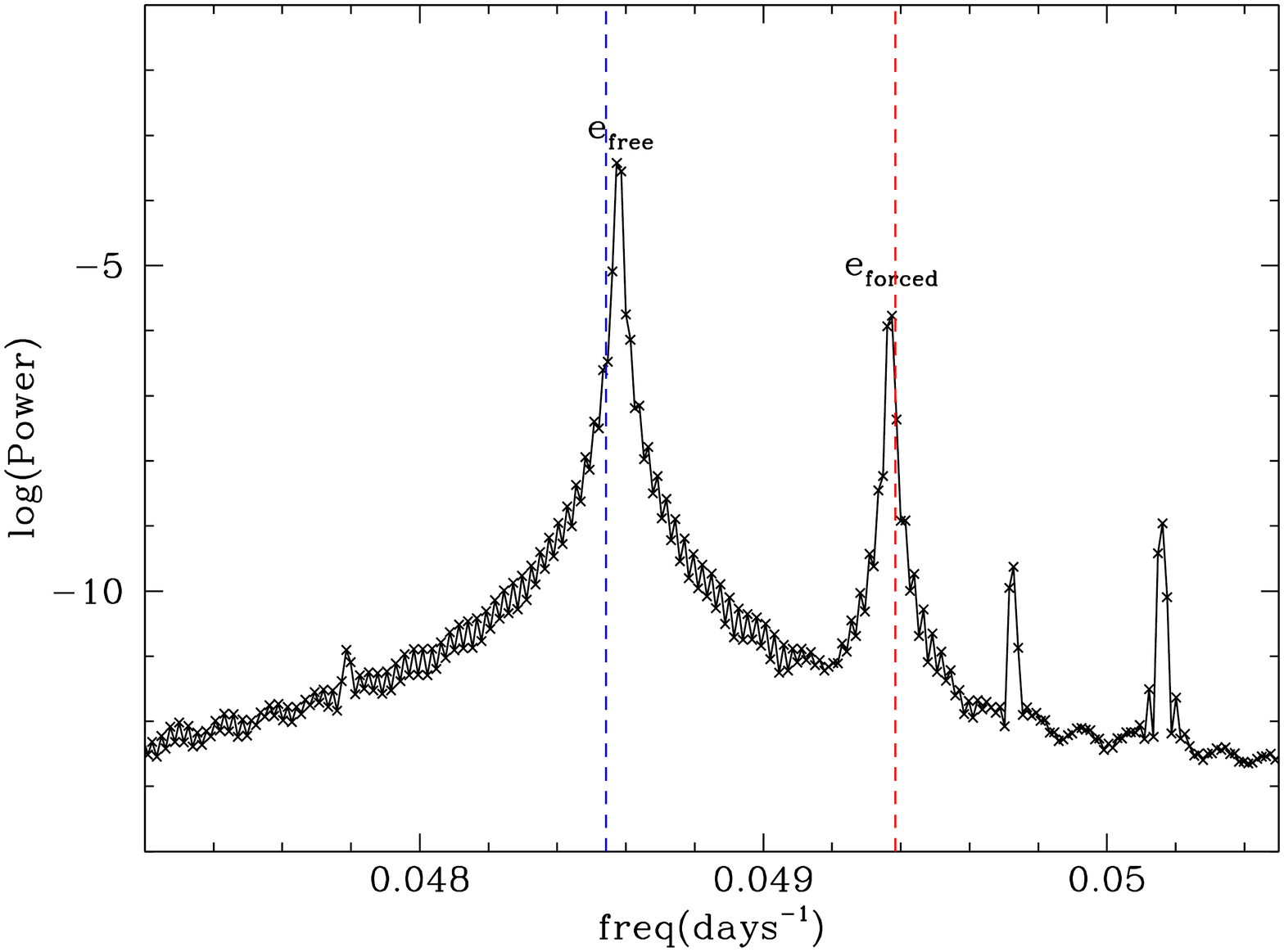}
\caption{Same as Figure \ref{spec_kep16}, but for Kepler-35 b.
\label{spec_kep35}
}
\end{center}
\end{figure}


\begin{figure}
\begin{center}
\includegraphics[angle=270,width=0.7\textwidth]{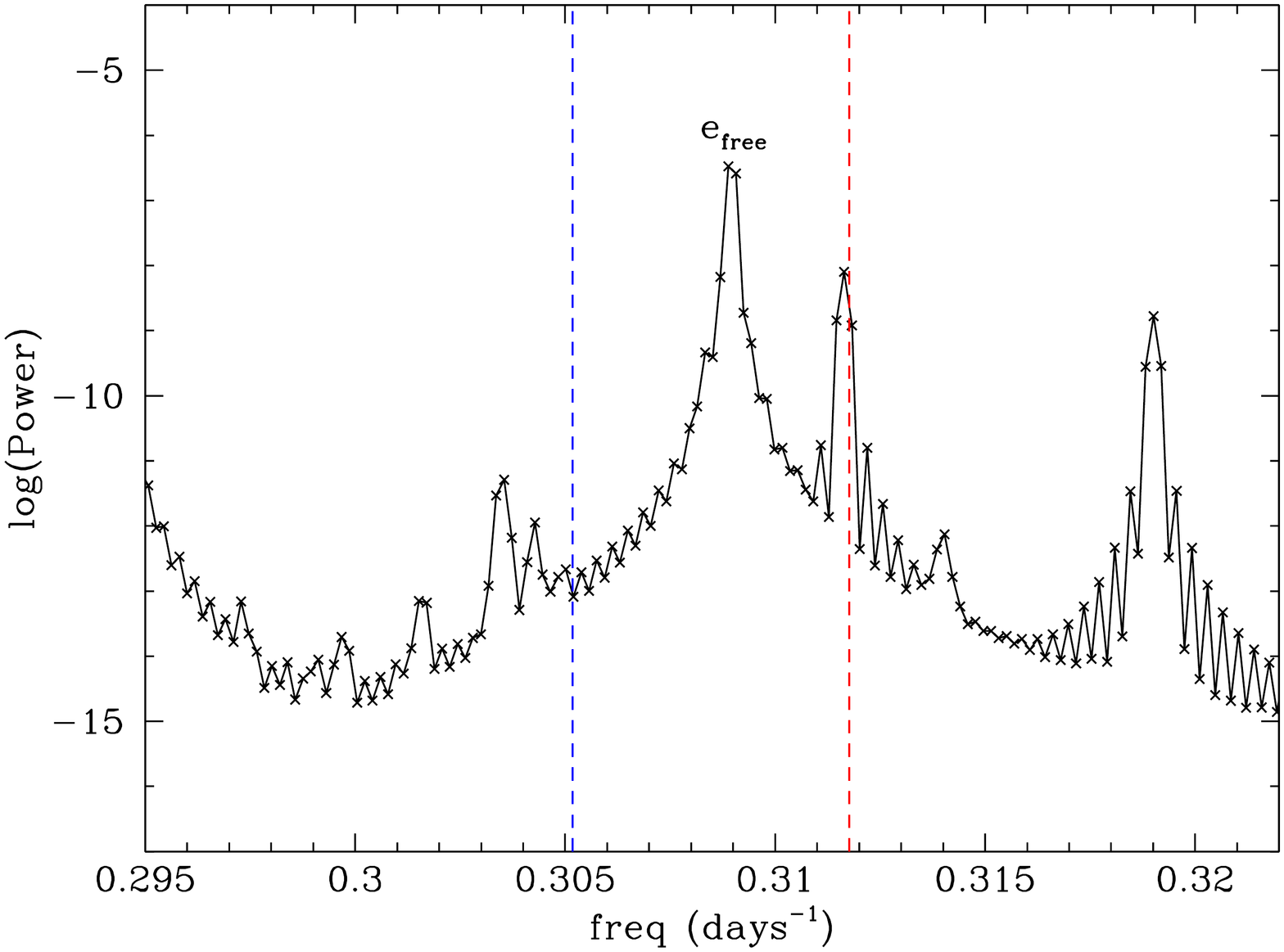}
\caption{Power spectrum obtained from FFT of $R(t)/R_0$ for Styx. See
Table \ref{table2} for the parameter values. The vertical blue and red
dashed lines indicate the analytic values of $\kappa_0$ and $n_0$,
respectively, from the first-order epicyclic theory.
\label{spec_styx}
}
\end{center}
\end{figure}

\begin{figure}
\begin{center}
\includegraphics[angle=270,width=0.7\textwidth]{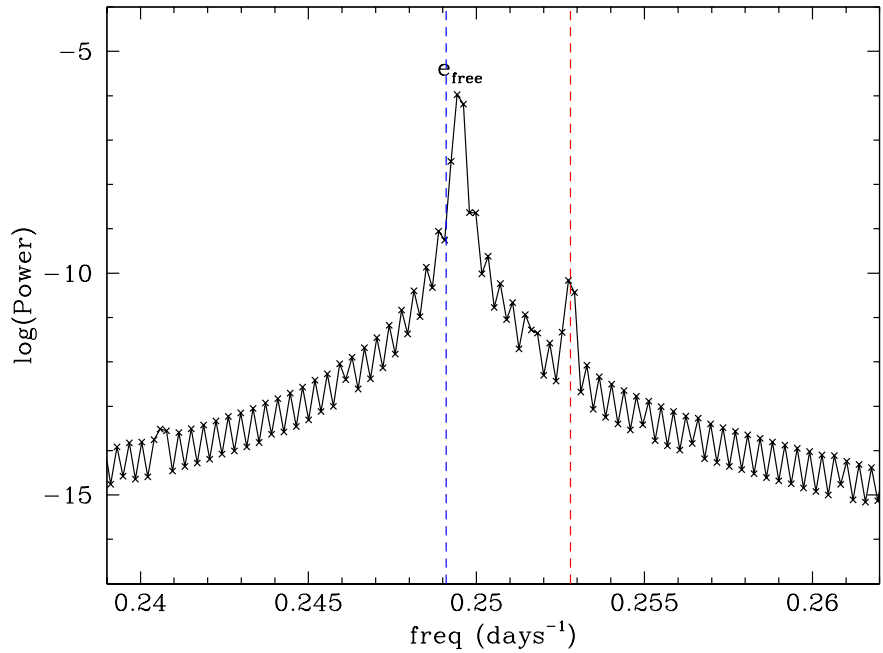}
\caption{Same as Figure \ref{spec_styx}, but for Nix.
\label{spec_nix}
}
\end{center}
\end{figure}


\begin{figure}
\begin{center}
\includegraphics[angle=270,width=0.7\textwidth]{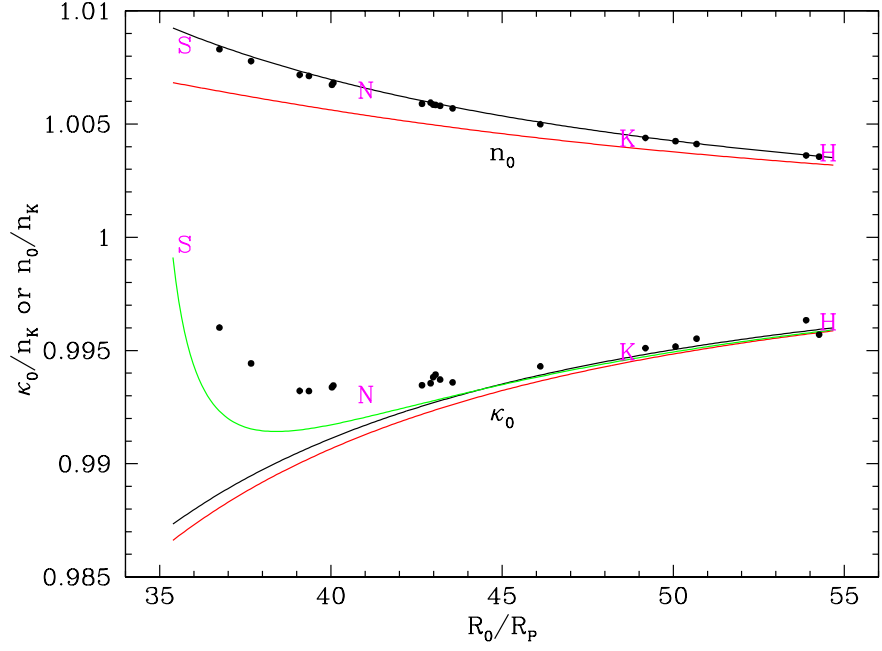}
\caption{Mean motion $n_0$ and epicyclic frequency $\kappa_0$ in units
of $n_K$ as a function of distance $R_0$ (in units of $R_P$) from the
centre of mass of Pluto-Charon.
Magenta alphabets and black dots show the numerical values from the
power spectrum for the satellites Styx, Nix, Kerberos, and Hydra and
the test particles from a simulation by \cite{woo18} (see text for
details).
Black, red, and green curves are from the first-order epicyclic
theory, the Hamiltonian theory with the secular term only, and the
Hamiltonian theory with the secular and 3:1 mean-motion resonance
terms, respectively.
\label{kappa_n}
}
\end{center}
\end{figure}


\begin{figure}
\begin{center}
\includegraphics[angle=270,width=0.7\textwidth]{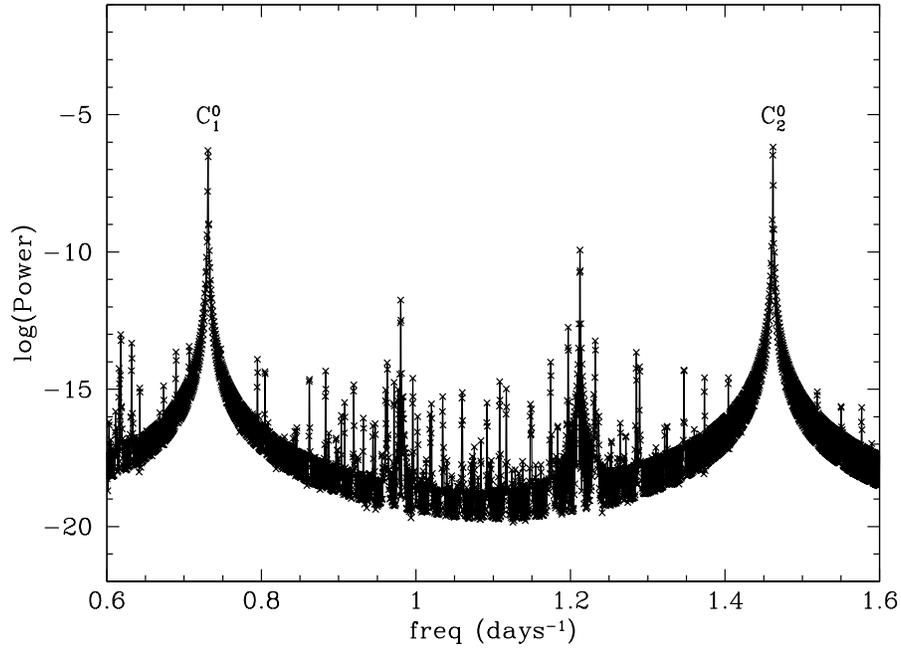}
\caption{Same as Figure \ref{spec_nix}, but with 2 times lower
  resolution and in a different frequency range showing the forced
  oscillation terms $|C_1^0| = 0.00126$ at $|n_0 - n_{AB}| =
  0.73091\,{\rm day}^{-1}$ and $|C_2^0| = 0.00142$ at $2|n_0 - n_{AB}|
  = 1.46182\,{\rm day}^{-1}$. 
\label{spec_nix_2}
}
\end{center}
\end{figure}


\begin{figure}
\begin{center}
\includegraphics[angle=270,width=0.7\textwidth]{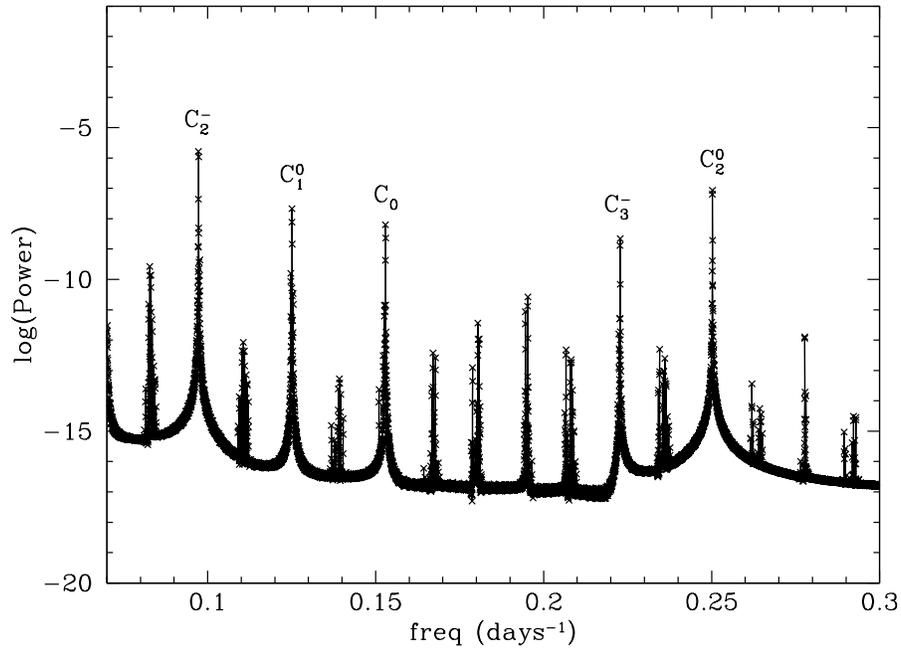}
\caption{Same as Figure \ref{spec_kep16},  but with 2 times lower
  resolution and in a different frequency range showing the five
  forced oscillation terms $C_1^0$, $C_2^0$, $C_0$, $C_2^-$ and $C_3^-$.
  See Table \ref{table3} for the comparison
  between the values of the forced oscillation terms from the power
  spectrum and the first-order epicyclic theory.
\label{spec_kep16_2}
}
\end{center}
\end{figure}

\end{document}